\documentclass[5p,11pt,sort&compress]{elsarticle}
\usepackage{graphicx}
\usepackage{color}
\usepackage{amssymb}
\usepackage{amsmath}
\usepackage[utf8]{inputenc}

\graphicspath{{./immagini/}} 
\journal{Physics Letters B}

\begin{document}

\begin{frontmatter}

\title{Polyakov loop and gluon quasiparticles: a self-consistent approach to Yang-Mills thermodynamics}

\author[unito]{Paolo Alba}
\author[unito]{Wanda Alberico}
\author[unito,unistate]{Marcus Bluhm}
\author[unica,infncat]{Vincenzo Greco}
\author[unito]{Claudia Ratti}
\author[unica]{Marco Ruggieri}

\address[unito]{Department of Physics, Torino University and INFN, Sezione di Torino, via P. Giuria 1, I-10125 Torino (Italy)}
\address[unica]{Department of Physics and Astronomy, University of Catania, Via S. Sofia 64, I-95123 Catania (Italy)}
\address[infncat]{INFN-Laboratori Nazionali del Sud, Via S. Sofia 62, I-95125 Catania (Italy)}
\address[unistate]{Department of Physics, North Carolina State University, Raleigh, NC 27965, USA}




\begin{abstract}
We present a quasiparticle model for the pure gauge sector of QCD, in which transverse quasigluons propagate in a Polyakov loop background field. By incorporating thermodynamic self-consistency in the approach, we show that our Polyakov loop extended quasiparticle model allows an accurate description of recent lattice results for all the thermodynamic quantities, including the Polyakov loop expectation value, in the deconfined phase. The related quasigluon mass exhibits a distinct temperature dependence, which is connected with the non-perturbative behavior seen in the scaled interaction measure of the pure gauge theory.
\end{abstract}
\end{frontmatter}


The possibility of creating the deconfined phase of QCD, the Quark-Gluon Plasma (QGP), in heavy-ion collision experiments at RHIC and LHC has triggered increasing theoretical activity, aimed at understanding the properties of nuclear matter under extreme conditions and the nature of its effective degrees of freedom~\cite{Jacak2012}. The unprecedented accuracy reached nowadays by lattice simulations of QCD 
thermodynamics~\cite{Borsanyi2010a,Borsanyi2010b,Cheng2010,Bazavov2012,Borsanyi2013} constitutes a further motivation to formulate effective models, whose predictions can be tested against the lattice results, thus allowing to validate or disprove the physics assumptions on which the models are based.

Interpreting the QGP in terms of quark and gluon quasiparticles, all the way down from the perturbative regime to the range of temperatures $T$ which can be reached in experiments, is an intriguing possibility which has received a lot of attention in recent years~\cite{Drago2004,Bluhm2007a,Bannur2008,Zhu2009,Buisseret2010,Chandra2011,Plumari2011,Cao2012,Das:2012ck}. The quasiparticle picture is based on the idea to view an interacting system as a system of weakly interacting quasiparticles which reflect important aspects of the interaction between the basic constituents of the system by effective properties as, for example, a temperature dependent mass. Such an approach was considered originally in~\cite{Biro1990,Peshier1996,Levai1998,Peshier2000} for describing deconfined, strongly interacting matter within a quasiparticle framework. Different approaches based on lattice~\cite{Silva:2013maa} as well as strong coupling expansion~\cite{Frasca:2010ce,Frasca2,Frasca:2007uz} have been proposed to investigate the 
spectrum 
of Yang-Mills theory; compared with these the simplicity of the quasiparticle concept makes it very attractive for a phenomenological treatment of hadronization, as well as for describing the QCD medium in the vicinity of the deconfinement transition temperature $T_c$. Recent lattice QCD results show that the transition is a smooth crossover \cite{Aoki2006}, which might indicate a survival of hadronic bound states or strong correlations between quarks above $T_c$ \cite{Ratti2012}. 

In order to establish which effective degrees of freedom populate the QGP in the vicinity of the deconfinement transition, it is important to clearly determine whether a quasiparticle formulation encounters limitations in its ability to describe thermodynamic quantities. For example it was noticed in~\cite{Plumari2011,Bluhm2005,Bluhm2007b} that, in its original form~\cite{Peshier1996,Peshier2000}, the quasiparticle model cannot reproduce with the same accuracy recent lattice QCD results for the pressure $p$, entropy density $s$ and energy density $\epsilon$, as well as quark number susceptibilities. Besides, the need to suppress the colored degrees of freedom in the vicinity of $T_c$ enforces unphysically large effective masses for quarks and gluons in this temperature region. 

Led by these motivations, the authors of Refs. \cite{Sasaki:2012bi,Ruggieri:2012ny} proposed a new version of quasiparticle models, being restricted to the pure gauge sector of QCD, in which transverse quasigluons propagate in a gluonic background field which is related to the Polyakov loop. The latter suppresses the colored degrees of freedom in the confined phase, thus removing the need for a large quasiparticle mass. Conceptually similar approaches have already been studied in~\cite{Meisinger2002,Meisinger2004}. 

In the present letter, we improve the work done in Ref.~\cite{Ruggieri:2012ny} by respecting thermodynamic self-consistency through the implementation of a temperature dependent bag function in line with~\cite{Gorenstein1995}. This allows not only to accurately reproduce continuum-extrapolated lattice gauge theory results for the thermodynamic quantities, but also a better description of the lattice results for the Polyakov loop. The pure-gauge results obtained in this work represent the basis on which we will build a thermodynamically self-consistent quasiparticle model for full QCD. A first attempt in this direction, which however neglects the inclusion of a $T$-dependent bag function, was recently reported in~\cite{Oliva2013}. 


In pure SU($N_c=$3) Yang-Mills thermodynamics, the only thermodynamic variable is the temperature. In case that the thermodynamic potential $\Omega$ depends not only explicitly on $T$ but also implicitly via phenomenological parameters $c_i(T)$, $\Omega(T,\{c_i(T)\})$ must satisfy additional consistency conditions reading~\cite{Gorenstein1995} 
\begin{equation}
\label{equ:statcondition}
 \left.\frac{\partial\Omega(T,\{c_i(T)\})}{\partial c_j}\right|_{T,\{c_i\}_{i\ne j}} = 0 \,. 
\end{equation}
These ensure thermodynamic self-consistency between the statistical mechanics definitions of the thermodynamic quantities. A prominent procedure used in order to fulfill the consistency conditions is to introduce a bag function $B(\{c_i(T)\})$ which depends on $T$ only implicitly via the phenomenological parameters $c_i(T)$. This function is often referred to as zero point energy density accounting for the fact that the lowest state energy of the system is now $T$-dependent~\cite{Gorenstein1995}. 

In this work, we combine the thermodynamics of the Polyakov loop with the thermodynamics of transverse gluon quasiparticles being coupled to the Polyakov loop and obeying a $T$-dependent dispersion relation of the form $\omega_g(k,T)=\sqrt{k^2+m_g^2(T)}$. Here, $k=|\vec{k}|$ is the momentum and $m_g(T)$ denotes the quasigluon mass, which enters the model as a phenomenological parameter such that the thermodynamic potential may be decomposed as 
\begin{eqnarray}
\nonumber
 \Omega(T,l_{\mathcal{F}}(T),m_g(T)) = \Omega_l(T,l_{\mathcal{F}}(T)) \hspace{6mm} & & \\
\label{equ:modelansatz}
 + \Omega_g(T,l_{\mathcal{F}}(T),m_g(T)) + B(m_g(T)) \,. & & 
\end{eqnarray}
This choice for $\Omega$ represents a minimalistic ansatz, which allows the construction of a thermodynamically self-consistent Polyakov loop extended quasiparticle approach. 

In Eq.~(\ref{equ:modelansatz}), the quantity $l_{\mathcal{F}}$ denotes the expectation value of the Polyakov loop~\cite{Polyakov1978} in the fundamental representation, which enters $\Omega$ in mean field approximation. The Polyakov loop $l_{\mathcal{R}}$ in the representation $\mathcal{R}$ of the gauge group SU($N_c$) follows as $l_{\mathcal{R}}={\rm Tr}_{\mathcal{R}}L_{\mathcal{R}}/d_{\mathcal{R}}$, where $d_{\mathcal{R}}$ is the dimension of the representation and the Polyakov line is defined as the Euclidean time ordered integral 
\begin{equation}
\label{equ:Polyline}
 L_{\mathcal{R}}(\vec{x})=\mathcal{T}\exp\left[ig\int_0^{1/T} d\tau\, A_4^{\mathcal{R}}(\vec{x},\tau)\right] \,. 
\end{equation}
Here, $g$ is the gauge coupling and $A_4^{\mathcal{R}}(\vec{x},\tau)=\sum_{a=1}^{N_c^2-1} A_4^a(\vec{x},\tau)\,t_a^{\mathcal{R}}$ denotes the vector potential in the time direction, where $t_a^{\mathcal{R}}$ are the generators of SU($N_c$) in representation $\mathcal{R}$.

The individual terms entering $\Omega$ in Eq.~(\ref{equ:modelansatz}) read as follows: for the pure Polyakov loop contribution $\Omega_l$ we make the ansatz 
\begin{equation}
\label{equ:Omega-l-contribution}
 \Omega_l(T,l_{\mathcal{F}}(T)) = a(T)f(l_{\mathcal{F}}(T)) \,,
\end{equation}
in which $a(T)$ summarizes the explicit temperature dependence of $\Omega_l$ such that $a(T)/T^4$ is dimensionless, while the function $f(l_{\mathcal{F}}(T))$ is defined as $f(l_{\mathcal{F}})=-5+6 l_{\mathcal{F}}^2+8 l_{\mathcal{F}}^3-9 l_{\mathcal{F}}^4$.
The form of the function $f(l_{\mathcal{F}})$ is a convenient choice, whose most important constraint is to preserve the $\mathcal{Z}_3$-symmetry.

The contribution stemming from the transverse gluon quasiparticles propagating in the Polyakov loop background field reads as 
\begin{equation}
\label{equ:Omega-g-contribution}
 \Omega_g=\frac{T}{\pi^2}\int_0^\infty dk\,k^2\,{\rm Tr}_{\mathcal{A}}\ln\left(1-L_{\mathcal{A}}e^{-\omega_g/T}\right)
\end{equation}
$L_{\mathcal{A}}$ being the Polyakov line Eq.~(\ref{equ:Polyline}) in the adjoint representation of the gauge group SU(3) while the trace is performed in color space. The ansatz in Eq.~(\ref{equ:Omega-g-contribution}) represents a phenomenologically motivated modification of the effective thermodynamic potential for the adjoint Polyakov loop at high temperatures~\cite{Weiss1981and1982}, in which the dispersion relation $\omega=k$ is replaced by hand through $\omega_g$. Assuming the relation $d_{\mathcal{A}}l_{\mathcal{A}}=d_{\mathcal{F}}^2l_{\mathcal{F}}^2-1$ with $d_{\mathcal{F}}=N_c$ and $d_{\mathcal{A}}=N_c^2-1$ to hold not only for the operators but also for the Polyakov loop expectation values, one may rewrite Eq.~(\ref{equ:Omega-g-contribution}) in terms of $l_{\mathcal{F}}$ as 
\begin{eqnarray}
\nonumber
 & & \!\!\!\!\!\!\!\!\!\!\!\!\!\!\!\!\!\!\!\!\!\! \Omega_g(T,l_{\mathcal{F}}(T),m_g(T)) = \frac{T}{\pi^2}\int_0^\infty dk\,k^2 \\
\nonumber
 & & \!\!\!\!\! \times\ln \bigg\{x^8\left(-1+x^{-3} +3 l_{\mathcal{F}} x^{-1}-3 l_{\mathcal{F}} x^{-2}\right)^2 \\
 & & \,\,\,\,\,\,\,\,\,\, \left(1+x^{-2}+x^{-1}(1+6 l_{\mathcal{F}}-9 l_{\mathcal{F}}^2)\right)\bigg\}
\end{eqnarray}
with $x=\exp\left[-\sqrt{(k^2+m_g^2)}/T\right]$. 


The expectation value of the Polyakov loop in the fundamental representation must satisfy the stationarity condition: 
\begin{eqnarray}
\nonumber
 0 & = & \left.\frac{\partial\Omega}{\partial l_{\mathcal{F}}}\right|_{T,m_g} \\
\label{equ:stat2new}
 & = & \left.\frac{\partial\Omega_l}{\partial l_{\mathcal{F}}}\right|_T + 
 \left.\frac{\partial\Omega_g}{\partial l_{\mathcal{F}}}\right|_{T,m_g}.
\end{eqnarray}
This condition implies 
\begin{equation}
\label{equ:afunction}
 \frac{a(T)}{T^4} = -\frac{1}{\pi^2 f'(l_{\mathcal{F}})} \int_0^\infty dy\, y^2 \,\mathcal{I}~,
\end{equation}
where
\begin{eqnarray}
\nonumber
 & & \!\!\!\!\!\!\!\!\!\!\!\!\!\!\! \mathcal{I} = 
 \frac{6-18l_{\mathcal{F}}}{1+6l_{\mathcal{F}}-9l_{\mathcal{F}}^2+2\,\text{Cosh}\left[\sqrt{y^2+m_g^2/T^2}\,\right]} \\
\label{equ:mathcalI}
 & & - \frac{6}{1-3l_{\mathcal{F}}+2\,\text{Cosh}\left[\sqrt{y^2+m_g^2/T^2}\,\right]} \,.
\end{eqnarray}

From the thermodynamic potential $\Omega$ in Eq.~(\ref{equ:modelansatz}), the equilibrium pressure 
$p(T,l_{\mathcal{F}}(T),m_g(T))$ is obtained as 
\begin{equation}
\label{equ:pressure}
 p = p_l + p_g - B 
\end{equation}
with $p_l=-\Omega_l$ and $p_g=-\Omega_g$. The other thermodynamic quantities follow from $p(T,l_{\mathcal{F}}(T),m_g(T))$ by using the thermodynamic identities $s=dp/dT$ and $\epsilon=Ts-p$, while the scaled interaction measure is defined as $I/T^4=(\epsilon-3p)/T^4$. 
In order to ensure the thermodynamical consistency we require that  
\begin{eqnarray}
\frac{dB}{dT}=\frac{dp_{g}}{dT}+\frac{dp_{l}}{dT}-\frac{\partial p_{g}}{\partial T}-\frac{\partial p_{l}}{\partial T},
\label{eq:pst}
\end{eqnarray}
which in turn implies that the total entropy is given by
\begin{equation}
\label{equ:entropydensity}
 s=s_l+s_g \,,
\end{equation}
as it can be easily verified. 
Thus the net effect of the bag function is to cancel the entropy density contribution which would arise from $dm_g/dT$.

In Eq.~(\ref{equ:entropydensity}), the terms read: 
\begin{eqnarray}
\nonumber
 & & \!\!\!\!\!\!\!\!\!\!\!\!\!\!\! s_l = \frac{2f(l_{\mathcal{F}})}{f'(l_{\mathcal{F}})} \frac{T^3}{\pi^2} \int_0^\infty dy\,y^2 \\
 & & \times\left(2 \,\mathcal{I} - 
 \frac{m_g^2}{T^2} \frac{{\rm Sinh}\left[\sqrt{y^2+m_g^2/T^2}\,\right]}{\sqrt{y^2+m_g^2/T^2}} \mathcal{H} \right) 
\end{eqnarray}
in which we have used Eq.~\eqref{equ:afunction} and $\mathcal{H}$ reads 
\begin{eqnarray}
\nonumber
 & & \!\!\!\!\!\!\!\!\!\!\!\!\!\!\!\!\!\!\!\!\!\!\!\!\! \mathcal{H} = \frac{6}{\left(1-3l_{\mathcal{F}}+2\,{\rm Cosh}\left[\sqrt{y^2+m_g^2/T^2}\,\right]\right)^2} \\
 & & \!\!\!\!\!\!\!\!\!\!\!\! + \frac{18l_{\mathcal{F}}-6}{\left(1+6l_{\mathcal{F}}-9l_{\mathcal{F}}^2+2\,{\rm Cosh}\left[\sqrt{y^2+m_g^2/T^2}\,\right]\right)^2} \,,\nonumber\\
 &&
\end{eqnarray}
while 
\begin{eqnarray}
\nonumber
 & & \!\!\!\!\!\!\!\!\!\!\!\!\!\!\!\!\!\!\!\!\! s_g = 
 -\frac{T^3}{\pi^2}\int_0^\infty dy\,\left(4y^2+\frac{m_g^2}{T^2}\right) \\ 
\nonumber
 & & \! \times\ln\Bigg[\tilde{x}^8\left(-1+\tilde{x}^{-3} +3l_{\mathcal{F}} \tilde{x}^{-1}-3l_{\mathcal{F}}\tilde{x}^{-2}\right)^2 \\
 & & \,\,\,\,\,\,\,\,\,\,\,\,\,\,\,\, \left(1+\tilde{x}^{-2}+\tilde{x}^{-1}\left[1+6l_{\mathcal{F}}-9l_{\mathcal{F}}^2\right]\right)\Bigg]
 \nonumber\\
 &&
\end{eqnarray}
with $\tilde{x}=\exp\left[-\sqrt{y^2+m_g^2/T^2}\,\right]$. 

From the above Eqs.~(\ref{equ:stat2new}) and~(\ref{equ:entropydensity}), the functions $a(T)$ and $m_g(T)/T$ can be determined unambiguously by adjusting $l_{\mathcal{F}}$ and the entropy density as functions of $T$ simultaneously to the first-principle lattice gauge theory results reported in~\cite{Lo2013} and~\cite{Borsanyi2012}, respectively. Moreover, the $T$-dependence of the function $B$ is most easily obtained from demanding that the pressure in Eq.~(\ref{equ:pressure}) also resembles the corresponding lattice data from~\cite{Borsanyi2012}.
We have verified numerically that the function $B(T)$ solves Eq. (\ref{eq:pst}), and so incorporates the effects of the $T$ dependence of $m_g(T)/T$ in the entropy density.

\begin{figure*}[t]
\begin{center}
\scalebox{.9}{
\includegraphics{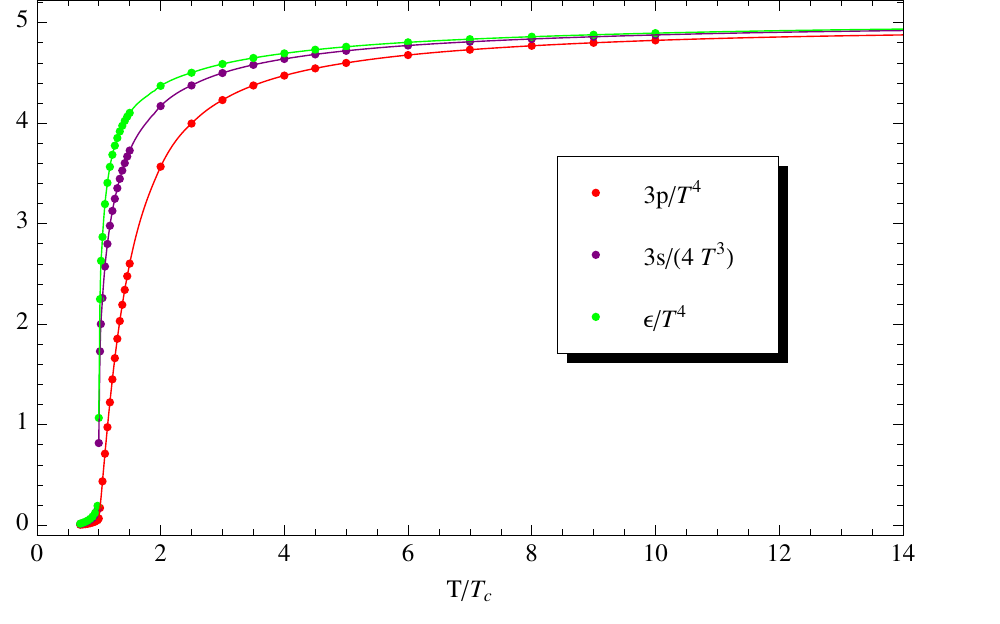}
\includegraphics{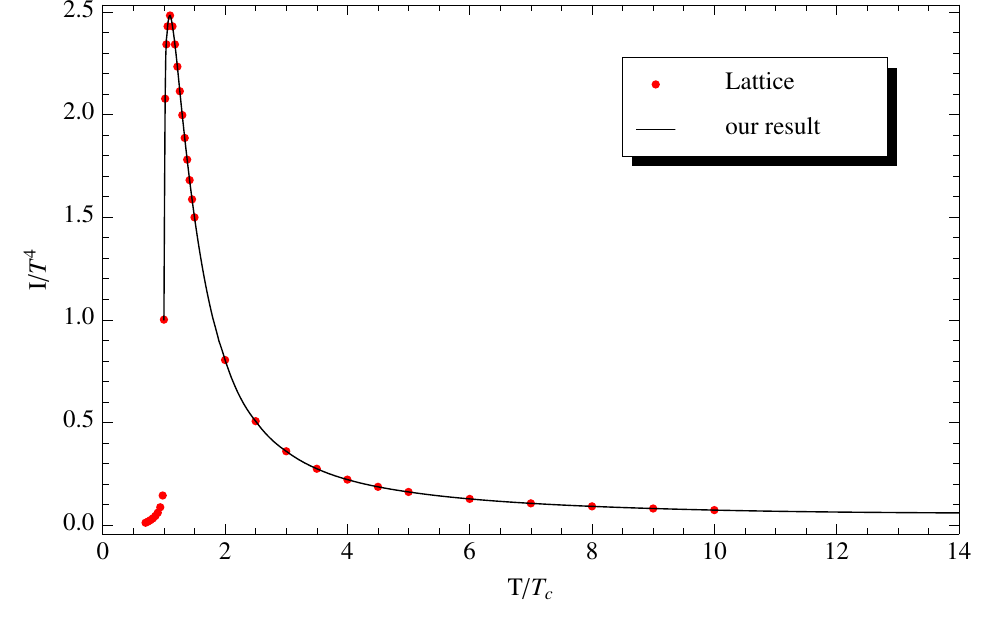}
}
\caption{(Color online) Left: scaled pressure $3p/T^4$, entropy density $3s/(4T^3)$ and energy density $\epsilon/T^4$ as functions of the scaled temperature $T/T_c$ obtained in our model (solid curves) and confronted with the corresponding lattice results (circles) from~\cite{Borsanyi2012}. Right: scaled interaction measure $I/T^4$ as a function of $T/T_c$.}
\label{fig1}
\end{center}
\end{figure*}
Our model results obtained in this way for the thermodynamic quantities $3p/T^4$, $3s/(4T^3)$, $\epsilon/T^4$ and $I/T^4$ as functions of $T/T_c$ are presented in Fig.~\ref{fig1} together with the corresponding 
lattice results. Here, we use a value of $T_c=270$~MeV~\cite{Kaczmarek2005}. 

\begin{figure}[t]
\begin{center}
\scalebox{.9}{
\includegraphics{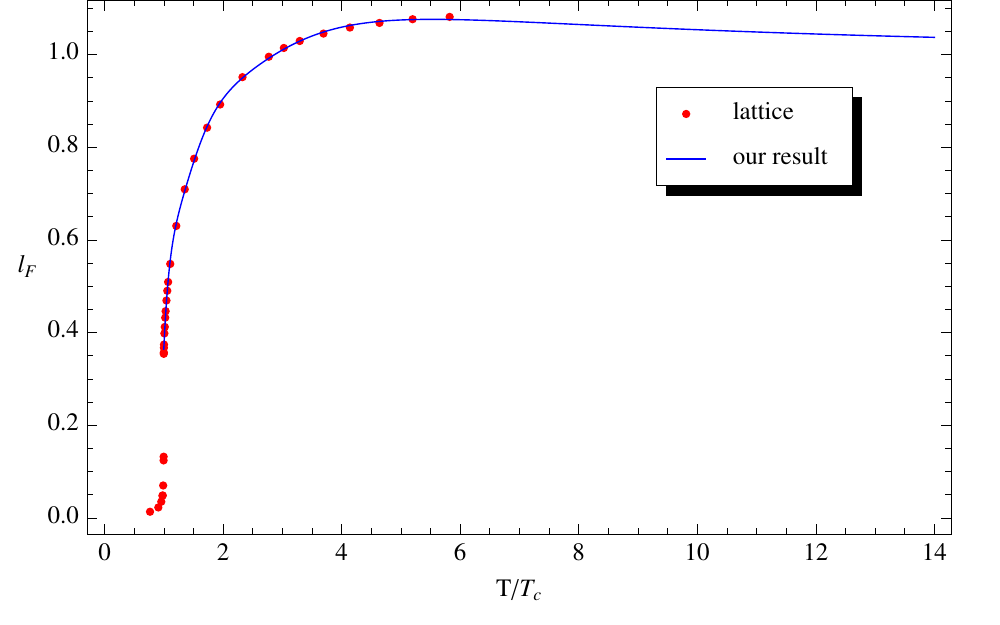}
}
\caption{(Color online) Expectation value of the Polyakov loop in the fundamental representation as a function of $T/T_c$. Our model results (solid curve) and the lattice results from~\cite{Lo2013} for the renormalized Polyakov loop (circles) are shown.}
\label{fig2}
\end{center}
\end{figure}

In Fig.~\ref{fig2}, we show the behavior of $l_{\mathcal{F}}$ as a function of the scaled temperature in comparison with the recent lattice results from~\cite{Lo2013}. As can be seen, the model approach is capable of accurately reproducing the behavior found in the lattice gauge theory calculations. We note, however, that while the lattice data increase monotonically in the evaluated temperature range, it is expected on general grounds from perturbative calculations~\cite{Gava1981} that in the asymptotic $T$-regime the Polyakov loop becomes a constant; notably $l_{\mathcal{F}}(T)\to 1$ is approached from above as $T\to\infty$, see also~\cite{Kaczmarek2005}. Such a behavior is incorporated in our approach. 

\begin{figure}[hb]
\begin{center}
\scalebox{.9}{
\includegraphics{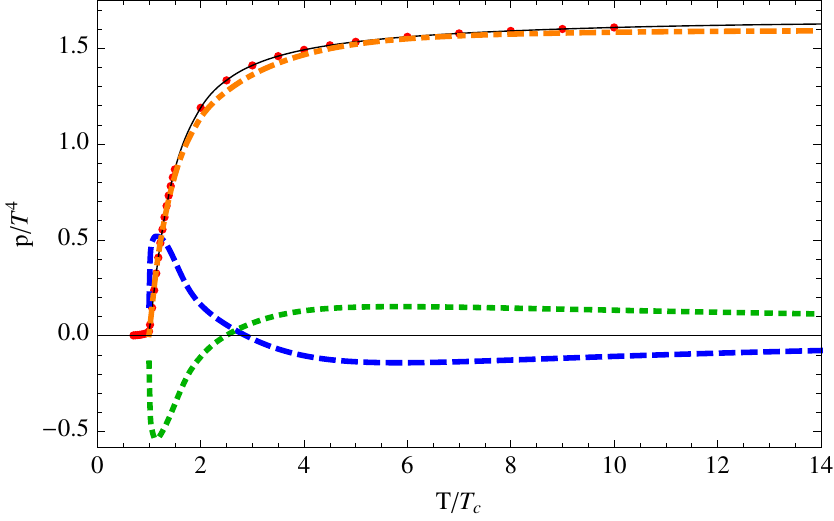}
}
\caption{(Color online) Individual contributions to the scaled pressure: the dashed, dash-dotted and dotted curves show the contributions stemming from $p_l/T^4$, $p_g/T^4$ and $-B/T^4$, respectively, which add up to the total scaled pressure $p/T^4$ (solid curve) in comparison with the lattice results from~\cite{Borsanyi2012} (circles).} 
\label{fig3}
\end{center}
\end{figure}

In Fig.~\ref{fig3}, we exhibit the individual contributions from Eq.~(\ref{equ:pressure}) to the scaled pressure $p/T^4$: as is evident, the pure Polyakov loop contribution $p_l/T^4$ is sizeable in the deconfinement transition region, while it approaches monotonically zero from below for large $T$ in line with $l_{\mathcal{F}}(T)\to 1^+$ for $T\to\infty$. The contribution $-B/T^4$ stemming from the bag function compensates $p_l/T^4$ to a large extent near $T_c$. This cancellation
largely stems from self-consistency requirements. Moreover, for large $T$ it gives important corrections to the transverse gluon quasiparticle component $p_g/T^4$, which starts to dominate the total scaled pressure above approximately $1.5\,T_c$. 

\begin{figure}[t]
\begin{center}
\scalebox{.9}{
\includegraphics{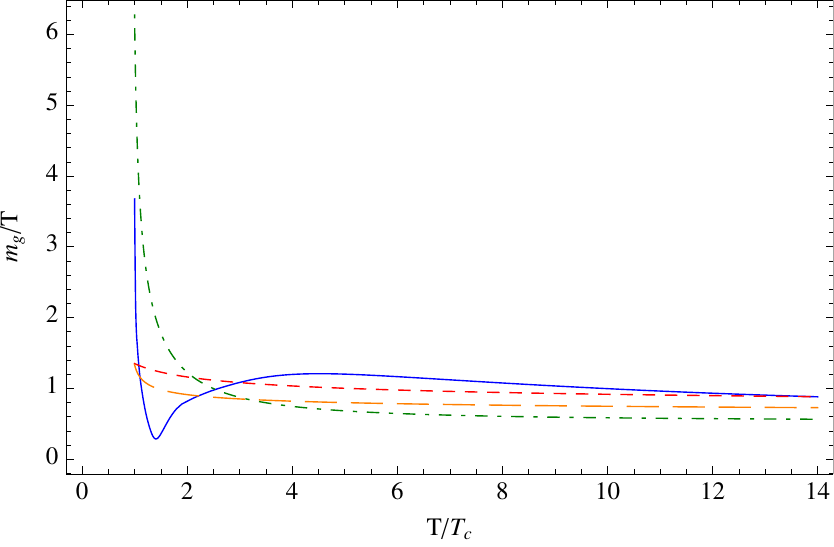}
}
\caption{(Color online) Scaled quasigluon mass $m_g(T)/T$ (solid curve) as a function of $T/T_c$. For comparison, the perturbative result $m_g(T)/T=\sqrt{g^2(T)/2}$ with $g^2(T)$ from Eq.~(\ref{equ:coupling}) for $c=1$, and the results from \cite{Ruggieri:2012ny,masscasto} are shown respectively by dashed, long-dashed and dash-dotted curves.} 
\label{fig4}
\end{center}
\end{figure}
The model results shown above for the thermodynamic quantities and the Polyakov loop expectation value $l_{\mathcal{F}}$ faithfully describe the available lattice data, because the function $m_g(T)/T$ was left undetermined initially. The scaled temperature dependence of this essential model ingredient is depicted in Fig.~\ref{fig4}. As can be seen, the scaled quasigluon mass $m_g(T)/T$ exhibits an interesting temperature dependence: for large $T$, one recovers the behavior known from perturbative calculations for the pure gauge sector of QCD, where at leading order $m_g(T)/T=\sqrt{g^2(T)/2}$ with 
\begin{equation}
\label{equ:coupling}
 g^2(T)=\left(\frac{11}{8\pi^2}\ln\left[2\pi c \frac{T}{T_c}\frac{T_c}{\Lambda_{\overline{\rm MS}}}\right]\right)^{-1}
\end{equation}
and $T_c/\Lambda_{\overline{\rm MS}}\simeq 1.14(4)$~\cite{Kaczmarek2004}. The parameter $c$ entering Eq.~(\ref{equ:coupling}) is usually considered in the range between $1/2$ and $2$. In Fig.~\ref{fig4}, we show for comparison the corresponding result for $c=1$. For $T\lesssim 4\,T_c$, however, non-perturbative effects become important and the $T$-dependence of $m_g(T)/T$ changes significantly. Most notably, the scaled quasigluon mass exhibits a pronounced minimum at about $1.4\,T_c$ with a minimal value of approximately $m_g\simeq 108$~MeV. 
This behavior is in striking contrast to the observations made in the previous approaches~\cite{Plumari2011,Peshier1996,Ruggieri:2012ny,Meisinger2004,Oliva2013}. As $T\to T_c^+$, the function $m_g(T)/T$ increases again drastically. A similar behavior is found in \cite{masscasto}, but we obtain a significative lower value of $m_g(T_c)\simeq 3.7\,T_c$; this value lies still within the ballpark of possible valence gluon masses for the lowest lying glueball states in line with the glueball pole mass values determined from analytical and lattice gauge theory estimates, see~\cite{Mathieu2009} for a recent review.
It is also interesting that, besides the small range of $T$ close to $T_c$, the function $m_g(T)/T$ found in the present calculation
is in quantitative agreement with a previous calculation based on a similar model with the Polyakov loop~\cite{Ruggieri:2012ny}.
The result of~\cite{Ruggieri:2012ny} is represented in Fig.~\ref{fig4} by the orange long-dashed curve, and it is clear that
the order of magnitude of the quasiparticle masses found within the two calculations is the same,
showing that in both models the Polyakov loop works as a suppressor of thermodynamical states permitting the quasiparticle
masses to be smaller than the ones obtained in the pure quasiparticle models.

\begin{figure}[ht]
\begin{center}
\scalebox{.9}{
\includegraphics{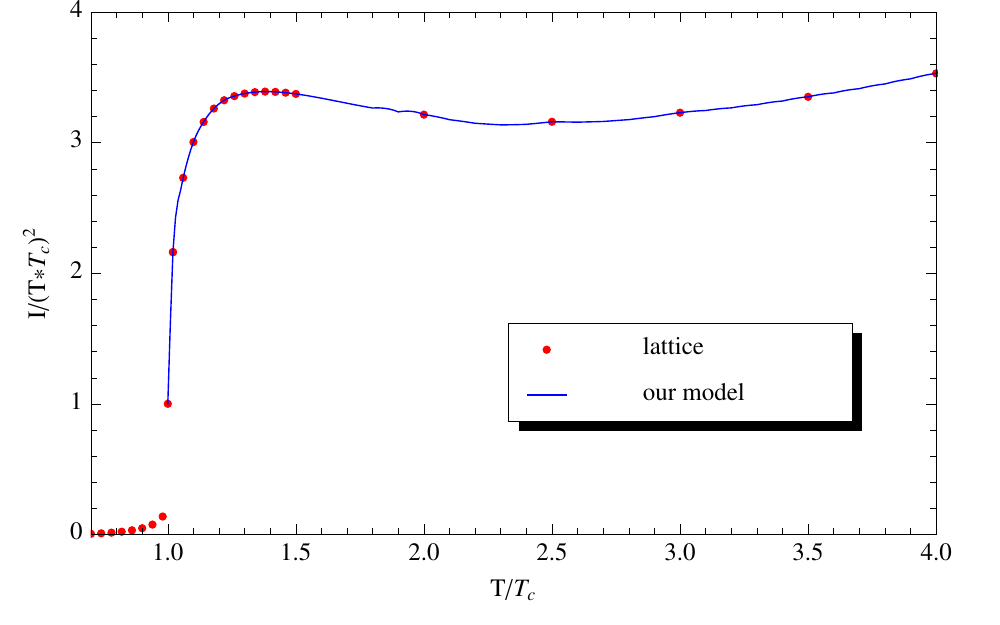}
}
\caption{(Color online) Scaled interaction measure $(I/T^4)\cdot(T/T_c)^2=(\epsilon-3p)/(T\!\cdot\!T_c)^2$ as a function of $T/T_c$ obtained in our model (solid curve) and compared with the lattice results (circles) from~\cite{Borsanyi2012}.} 
\label{fig5}
\end{center}
\end{figure}

In order to discuss the onset of a non-perturbative behavior in the thermodynamic quantities, it was suggested in~\cite{Pisarski2007} to consider $(I/T^4)\cdot(T/T_c)^2$ rather than $I/T^4$. In this way, deviations from the perturbatively expected $T^4$-dependence (up to logarithmic corrections) of the interaction measure can be highlighted. The lattice calculations reported in~\cite{Borsanyi2012} identified indeed a $T^2$-dependence in $(\epsilon-3p)$ as the dominant non-perturbative effect in the deconfined phase for temperatures up to approximately $4\,T_c$. Various models were shown to successfully explain such a $T^2$-behavior, at least qualitatively~\cite{Pisarski2007,Pisarski2006,Castorina2011,Dumitru2012,Sasaki2013}. In our approach, the $T^2$-dependence in the interaction measure arises naturally from the presence of massive quasigluons in the model~\cite{Meisinger2002}. In Fig.~\ref{fig5}, we show $I/T^4$ multiplied by $(T/T_c)^2$ as a function of $T/T_c$ in 
comparison with the lattice results from~\cite{Borsanyi2012}. In agreement with the latter, we find little sensitivity of $(\epsilon-3p)/(T\!\cdot\!T_c)^2$ on $T$ in the range $1.2<T/T_c<4$, suggesting a dominating $T^2$-behavior in the interaction measure stemming from non-perturbative effects in this temperature regime. 

In conclusion, we presented a thermodynamically self-consistent quasiparticle model for pure $SU(3)$ Yang-Mills thermodynamics, in which transverse gluon quasiparticles with a temperature dependent mass propagate in a Polyakov loop background field. We have shown that within our approach it is possible to accurately describe the recent lattice results from~\cite{Lo2013,Borsanyi2012} for all the thermodynamic quantities, including the Polyakov loop expectation value, in the deconfined phase. We restricted ourselves to $T\geq T_c$, while recent hybrid approaches~\cite{Sasaki:2012bi,Sasaki2013} aim at a combined picture for both confined and deconfined phases. 

In line with the successful description of the available lattice results, the quasigluon mass shows a distinct temperature dependence, which is connected with the non-perturbative behavior observed in the scaled interaction measure. In particular, the development of the pronounced minimum in $m_g(T)/T$ just above $T_c$ points towards interesting dynamical consequences, which might be explored in fourthcoming studies. The model we discussed here will serve as a basis for a quasiparticle description of the thermodynamics of QCD including dynamical quarks. Work in this direction is in progress.

\section*{Acknoledgements}
This work is partially supported by funds provided by the Italian Ministry of Education, Universities and Research
under the Firb Research Grant RBFR0814TT; V.G. acknowledges the ERC-STG funding under the QGPDyn grant.\\


\end{document}